\documentclass[12pt,a4paper,twoside]{article}
\usepackage{amsmath,amsfonts,amsthm,graphicx}
\numberwithin{equation}{section}

\begin{document}
\title{
From the quantum Jacobi-Trudi and Giambelli formula to 
 a nonlinear integral equation for 
thermodynamics of 
 the higher spin Heisenberg model 
}
\author{ 
Zengo Tsuboi
\footnote{E-mail address: tsuboi@issp.u-tokyo.ac.jp}
\\
{\it Institute for Solid State Physics, 
University of Tokyo,} \\ 
{\it Kashiwanoha 5-1-5, Kashiwa, Chiba, 277-8581, Japan}
}
\date{}
\maketitle
\begin{abstract}
We propose a nonlinear integral equation (NLIE) with only 
one unknown function, 
which gives the free energy of the integrable one dimensional
 Heisenberg 
model with arbitrary spin. In deriving the NLIE, 
the quantum Jacobi-Trudi and Giambelli formula 
(Bazhanov-Reshetikhin formula), 
which gives the solution of the $T$-system, plays an important role. 
In addition, we also calculate the high temperature 
expansion of the specific heat and the magnetic susceptibility. 
\end{abstract}
Short title: Nonlinear integral equation \\
{\it MSC:} 82B23; 45G15; 82B20; 17B80 \\
{\it PACS2001:} 02.30.Rz; 02.30.Ik; 05.50.+q; 05.70.-a \\
{\it Key words:}
nonlinear integral equation; 
higher spin Heisenberg model; 
quantum transfer matrix; 
Jacobi-Trudi and Giambelli formula; 
thermodynamic Bethe ansatz; 
$T$-system \\\\
\noindent to appear in J. Phys.A: Math. Gen.
%
\section{Introduction}
Thermodynamic Bethe ansatz (TBA) equations have been used to 
analyze thermodynamics of various kind of solvable lattice models \cite{Ta99}. 
However it is not always easy to treat TBA equations 
since in general they contain an infinite number of unknown 
functions. 
So there are several attempts to simplify the TBA equations. 
In particular for the 1D spin $\frac{1}{2}$ Heisenberg model, 
Kl\"umper proposed nonlinear integral equations (NLIE) \cite{K93} 
which contain a finite number of unknown functions 
from the point of view of the quantum transfer matrix (QTM) 
approach \cite{S85,SI87,K87,SAW90,K92,K93}. 
There is also a similar NLIE by Destri and de Vega \cite{DD92}. 

Another type of NLIE for the spin $\frac{1}{2}$ $XXZ$ model, 
which contains only one unknown function, was 
discovered by Takahashi \cite{Ta01} to simplify the TBA equation. 
Later, this NLIE was rederived \cite{TSK01} from the $T$-system \cite{KR87,KP92} 
of the QTM. 
Moreover this NLIE for the spin $\frac{1}{2}$ $XXX$ model was also rederived 
from a fugacity expansion formula \cite{KW02}. 
We derived this type of NLIE for the $osp(1|2s)$ model \cite{T02} 
and the $sl(r+1)$ Uimin-Sutherland model \cite{T03} 
for {\em arbitrary} rank. The number of the unknown functions of these NLIE 
coincides with the rank of the underlying algebras. 
All of them are NLIE for fundamental representations of underling algebras. 
So it is desirable to derive NLIE for tensor representations, i.e. 
NLIE for higher spin models. 
The purpose of this paper is to derive NLIE for the Heisenberg spin 
chain with {\em arbitrary} spin $\frac{s}{2}$ \cite{B83}. 

Thermodynamics of the higher spin Heisenberg model 
was firstly investigated \cite{B83} by the TBA equations, which 
contain an infinite number of unknown functions. 
Later a set of NLIE with $s+1$ unknown functions for this model 
was derived \cite{S99} by the QTM approach. 
This NLIE is an extension of the Kl\"umper's type of NLIE \cite{K93}. 
On the other hand, our new NLIE contains only {\em one} unknown function 
for {\em arbitrary} $s$. Thus, as far as the number of the unknown 
functions, our new NLIE is a further simplification of the TBA equations. 

In section 2, we introduce the higher spin Heisenberg model 
and define the $T$-system of the QTM. 
As a solution of the $T$-system, we introduce 
the quantum Jacobi-Trudi and Giambelli formula 
(Bazhanov-Reshetikhin formula \cite{BR90}) (\ref{BR}), 
which plays an essential role in the derivation of the NLIE. 
This formula expresses an eigenvalue formula of the transfer matrix 
 for the tensor representation in the auxiliary space 
in a determinant form. 
In the representation theoretical context, it may be viewed 
as a Yangian analogue of classical Jacobi-Trudi and Giambelli formula 
on Schur functions \cite{KOS95,BR90}. 
In section 3, we derive the NLIE (\ref{nlie4}), which is our main result. 
The $T$-system which we have to use here 
is not the standard one \cite{KP92} (\ref{t-sys}) but 
an old one \cite{KR87} (\ref{t-sys2}), (\ref{t-sys3}). 
Due to the quantum Jacobi-Trudi and Giambelli formula, 
determinants appear in our new NLIE. 
These novel situations contrast with 
the fundamental representation cases \cite{Ta01,TSK01,T02,T03}. 
Using our new NLIE (\ref{nlie4}), we also 
calculate the high temperature expansion of the 
specific heat and the magnetic susceptibility in section 4. 
It will be difficult to obtain the same result by 
the traditional TBA equations. 
Section 5 is devoted to concluding remarks. 
\section{QTM method, $T$-system, 
quantum Jacobi-Trudi and Giambelli formula}
We introduce the higher spin Heisenberg model, and 
define the QTM, the $T$-system and  the quantum Jacobi-Trudi 
and Giambelli formula for this model. 
A more detailed explanation of the QTM analysis for this model 
can be found in \cite{S99}. 

The Hamiltonian of the spin $\frac{s}{2}$ Heisenberg model 
is given as follows \cite{B83}.
\begin{eqnarray}
{\mathcal H}_{0}=J \sum_{j=1}^{L} 
{\mathcal Q}_{s}({\mathbf S}_{j}{\mathbf S}_{j+1}),
\label{hamiltonian}
\end{eqnarray}
where ${\mathbf S}_{j}=({S}_{j}^{x},{S}_{j}^{y},{S}_{j}^{z})$ 
is the spin $\frac{s}{2}$ operator 
acting on the $j$-th site, and ${\mathbf S}_{j}{\mathbf S}_{j+1}=
S_{j}^{x}S_{j+1}^{x}+S_{j}^{y}S_{j+1}^{y}+S_{j}^{z}S_{j+1}^{z}$.  
$J$ is a real coupling constant ($J>0$ (resp. $J<0$) 
corresponds to the anti-ferromagnetic (resp. ferromagnetic) regime)
 and $L$ is the number of the lattice sites. 
We assume the periodic boundary condition ${\mathbf S}_{L+1}={\mathbf S}_{1}$. 
${\mathcal Q}_{s}(x)$ is defined as 
\begin{eqnarray}
{\mathcal Q}_{s}(x)=-2\sum_{j=0}^{s-1}\sum_{k=j+1}^{s}\frac{1}{k}
 \prod_{p=0(p \ne j)}^{s} \frac{x-x_{p}}{x_{j}-x_{p}},
\end{eqnarray}
where $x_{p}=\frac{1}{2}(p(p+1)-s(\frac{s}{2}+1))$. 
For $s=1$ and $s=2$ case, (\ref{hamiltonian}) 
becomes 
\begin{eqnarray}
{\mathcal H}_{0}=
2J \sum_{j=1}^{L}\{{\mathbf S}_{j}{\mathbf S}_{j+1}-\frac{1}{4} \},
\quad 
{\mathcal H}_{0}=\frac{J}{2} \sum_{j=1}^{L}\{{\mathbf S}_{j}{\mathbf S}_{j+1}- 
({\mathbf S}_{j}{\mathbf S}_{j+1})^2 \},
\end{eqnarray}
respectively. 
The $R$-matrix \cite{KRS81,SAA83} of 
the classical counter part of the spin $\frac{s}{2}$ Heisenberg model is 
defined as 
\begin{eqnarray}
R(v)=\sum_{j=0}^{s} (\prod_{k=j+1}^{s}\frac{v-k}{v+k}) P^{j},
\label{rmat}
\end{eqnarray}
where $P^{j}$ is the projector onto $j+1$ dimensional irreducible 
$sl(2)$ module. $P^{j}$ can be expressed as 
\begin{eqnarray}
P^{j}=\prod_{p=0(p \ne j)}^{s}
\frac{{\mathbf S}\otimes {\mathbf S}-x_{p}}{x_{j}-x_{p}},
\end{eqnarray}
where ${\mathbf S}=(S^{x},S^{y},S^{z})$ 
is the spin $\frac{s}{2}$ operator, 
and ${\mathbf S}\otimes
 {\mathbf S}=S^{x}\otimes S^{x}+S^{y}\otimes S^{y}+S^{z}\otimes S^{z}$. 
In our normalization of the $R$-matrix,  
$R(\infty)$ is an identity operator. 
The row to row transfer matrix is defined as 
\begin{eqnarray}
t(v)={\mathrm tr}_{0}R_{0L}(v)\cdots R_{02}(v)R_{01}(v),
\end{eqnarray}
where $R_{0j}(v)$ is the $R$-matrix (\ref{rmat}) 
acting on the auxiliary space and the $j$-th site of 
the quantum space. This transfer matrix is connected to the 
hamiltonian (\ref{hamiltonian}) as
\begin{eqnarray}
{\mathcal H}_{0}=J\frac{d}{dv} \log t(v)|_{v=0}.
\end{eqnarray}
One can add an external field term 
${\mathcal H}_{ex}=-2h\sum_{j}S_{j}^{z}$ 
in the hamiltonian (\ref{hamiltonian})
without breaking the integrability as 
${\mathcal H}={\mathcal H}_{0}+{\mathcal H}_{ex}$. 
The QTM is defined 
as
\begin{eqnarray}
t_{{\rm QTM}}(v)={\mathrm tr}_{0} e^{\frac{2hS_{0}^{z}}{T}}
\widetilde{R}_{N0}(u-iv)R_{N-1 \ 0}(u+iv)\cdots 
\widetilde{R}_{20}(u-iv)R_{10}(u+iv),
\label{QTM}
\end{eqnarray}
where $\widetilde{R}_{j k}(v)$ is 
defined by \symbol{96}$90^{\circ}$ rotation\symbol{39} of $R_{kj}(v)$, i.e.,
$\widetilde{R}_{j k}(v)={}^{t_k} \! R_{k j}(v)$ ($t_k$:
the transposition of $R_{kj}(v)$ in the $k$-th space); 
$N$ is the Trotter number and assumed to be even; 
$u=-\frac{J}{T N}$ ($T$ is a temperature); 
the Boltzmann constant is set to $1$. 
The free energy per site is expressed 
in terms of  the largest eigenvalue $\widetilde{T}_{s}(v)$ of 
the QTM (\ref{QTM}) at $v=0$:
\begin{eqnarray}
f=
-T\lim_{N\to \infty}\log \widetilde{T}_{s}(0).
\label{free}
\end{eqnarray} 
We can embed $\widetilde{T}_{s}(v)$ into the 
eigenvalue formulae $\{\widetilde{T}_{m}(v)\}_{m \in {\mathbb Z}_{\ge 0}}$
 for the fusion hierarchy of the QTM, i.e. a Yang-Baxterization of the 
 character of the $m$-th symmetric tensor representation of $sl(2)$:
\begin{eqnarray}
\widetilde{ T}_{m}(v)=T_{m}(v)/{\mathcal N}_{m}(v),
\label{DVF}
\end{eqnarray}
where 
\begin{eqnarray}
&& \hspace{-20pt}
T_{m}(v)=\sum_{1\le d_{1}\le d_{2} \le \cdots \le d_{m} \le 2}
\prod_{j=1}^{m}z(d_{j} ;v+\frac{i}{2}(m-2j+1))
\nonumber \\
&&=\sum_{k=0}^{m}e^{\frac{(m-k)\mu_{1}+k\mu_{2}}{T}}
\prod_{j=1}^{m-k}\phi_{-}(v+\frac{m-s-2j}{2}i)\phi_{+}(v+\frac{m-s-2j+2}{2}i)
\nonumber \\
&& \times 
\prod_{j=m-k+1}^{m}\phi_{-}(v+\frac{m+s-2j}{2}i)\phi_{+}(v+\frac{m+s-2j+2}{2}i)
\nonumber \\
&& \times 
\frac{Q(v+\frac{m}{2}i)Q(v-\frac{m+2}{2}i)}
{Q(v-\frac{m-2k}{2}i)Q(v-\frac{m-2k+2}{2}i)};
\label{DVF2}
\end{eqnarray}
\begin{eqnarray}
{\mathcal N}_{m}(v)=\prod_{k=1}^{m}
 \phi_{-}(v+\frac{m-2k-s}{2}i)\phi_{+}(v+\frac{m-2k+s+2}{2}i);
\end{eqnarray}
 $\phi_{\pm}(v)=(v \pm iu)^{\frac{N}{2}}$; 
 $Q(v)=\prod_{j=1}^{M}(v-v_{j})$ ($M \in {\mathbb Z}_{\ge 0}$);
  $\mu_{1}$ and $\mu_{2}$ are chemical potentials 
(in our case, $\mu_{1}=h$, $\mu_{2}=-h$); 
\begin{eqnarray}
&& z(1;v)=e^{\frac{\mu_{1}}{T}} 
\phi_{-}(v-\frac{s+1}{2}i)\phi_{+}(v-\frac{s-1}{2}i)
  \frac{Q(v+\frac{i}{2})}{Q(v-\frac{i}{2})},
\nonumber \\ 
&& z(2;v)=e^{\frac{\mu_{2}}{T}} \phi_{-}(v+\frac{s-1}{2}i)
\phi_{+}(v+\frac{s+1}{2}i)
  \frac{Q(v-\frac{3i}{2})}{Q(v-\frac{i}{2})}.
\end{eqnarray}
 Here we adopt a convention 
 $\prod_{j=1}^{0}(\cdots)=\prod_{j=m+1}^{m}(\cdots)=1$. 
 $v_{j}$ is a solution of the Bethe ansatz equation
\begin{eqnarray}
-\frac{\phi_{-}(v_{k}-\frac{si}{2})\phi_{+}(v_{k}-\frac{si}{2}+\frac{gi}{2})}
{\phi_{-}(v_{k}+\frac{si}{2})\phi_{+}(v_{k}+\frac{si}{2}+\frac{gi}{2})}
=e^{-\frac{\mu_{1}-\mu_{2}}{T}} \frac{Q(v_{k}-i)}{Q(v_{k}+i)},
\ k \in \{1,2,\dots, M \},
\label{BAE}
\end{eqnarray}
where $g=2$: the dual Coxeter number. 
Note that 
$T_{m}(v)$ (\ref{DVF2}) is free of poles under the 
Bethe ansatz equation (\ref{BAE}). 
One can show that the function $\widetilde{ T}_{m}(v)$ (\ref{DVF}) 
satisfies the following $T$-system \cite{KP92}. 
\begin{eqnarray}
\widetilde{T}_{m}(v-\frac{i}{2})\widetilde{T}_{m}(v+\frac{i}{2})
=
\widetilde{T}_{m-1}(v)\widetilde{T}_{m+1}(v)
+\widetilde{g}_{m}(v), 
\quad m \in {\mathbb Z}_{\ge 1},
\label{t-sys}
\end{eqnarray}
where 
\begin{eqnarray}
&& \widetilde{ T}_{0}(v)=1, \\
&& \widetilde{g}_{m}(v)=e^{\frac{m(\mu_{1}+\mu_{2})}{T}}
\prod_{k=1}^{\min(m,s)}
 \frac{\phi_{-}(v+\frac{m+s+1-2k}{2}i)\phi_{+}(v-\frac{m+s+1-2k}{2}i)}
 {\phi_{-}(v-\frac{m+s+1-2k}{2}i)\phi_{+}(v+\frac{m+s+1-2k}{2}i)}.
\end{eqnarray}
The solution of the $T$-system (\ref{t-sys}) is given by the following 
quantum Jacobi-Trudi and Giambelli 
formula (Bazhanov-Reshetikhin formula \cite{BR90}). 
\begin{eqnarray}
\widetilde{ T}_{m}(v)= \det _{1\le j,k \le m}
\left({\widetilde f}^{1+j-k}
\left(
v-\frac{j+k-m-1}{2}i
\right) 
\right), \label{BR}
\end{eqnarray}
where the matrix elements are given as follows
\begin{eqnarray}
{\widetilde f}^{a}(v)=
\left\{ 
\begin{array}{cl}
 1 & a=0 \\
 {\tilde T}_{1}(v) & a=1 \\
 {\tilde g}_{1}(v) & a=2 \\ 
 0 & a >2 \quad {\rm or} \quad a<0 .
 \end{array}
 \right.
\end{eqnarray}
The following functional relations \cite{KR87} are equivalent to the 
$T$-system (\ref{t-sys}). 
\begin{eqnarray}
&& \hspace{-50pt}
\widetilde{ T}_{1}(v)\widetilde{ T}_{m-1}(v-\frac{m}{2}i)
=
\widetilde{ T}_{m}(v-\frac{m-1}{2}i)
+\widetilde{g}_{1}(v-\frac{i}{2})
\widetilde{ T}_{m-2}(v-\frac{m+1}{2}i),  \label{t-sys2} \\
&& \hspace{-50pt} 
\widetilde{ T}_{1}(v)\widetilde{ T}_{m-1}(v+\frac{m}{2}i)
=\widetilde{ T}_{m}(v+\frac{m-1}{2}i)
+\widetilde{g}_{1}(v+\frac{i}{2})
\widetilde{ T}_{m-2}(v+\frac{m+1}{2}i),  \label{t-sys3} \\
&& \hspace{120pt}\quad m \in {\mathbb Z}_{\ge 1}.
\nonumber 
\end{eqnarray}
(\ref{t-sys2}) follows from an expansion of 
the determinant (\ref{BR}) down the first row. 
(\ref{t-sys3}) follows from an expansion of 
the determinant (\ref{BR}) down the last column. 
\section{A new nonlinear integral equation}
Suzuki had \cite{S99} the following observation 
from numerics. 

{\it For the largest eigenvalue sector of $\widetilde{ T}_{s}(v)$, 
the corresponding root of the BAE (\ref{BAE}) forms $s$-strings. 
In this case, imaginary parts of the 
zeros $\{\tilde{z}_{m}\}$ of $\widetilde{ T}_{m}(v)$ ($m=1,2,\dots,s$) 
are located 
near the lines ${\mathrm Im}v=\pm \frac{1}{2}(s+m-2j)$, 
$j=0,1,\dots,m-1$. }

\begin{figure}
\begin{center}
\includegraphics[width=0.95\textwidth]
{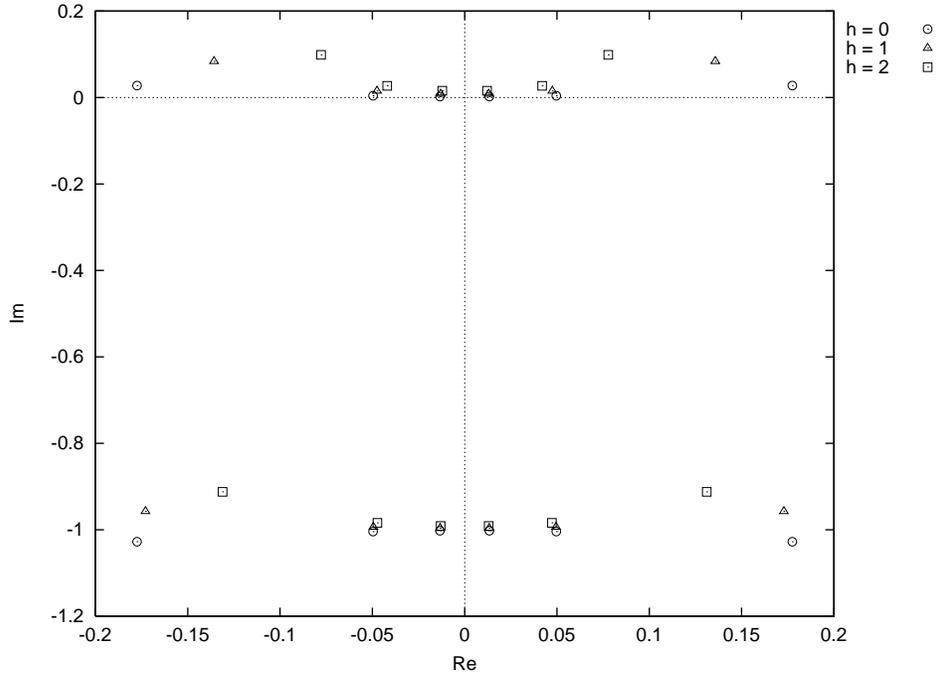}
\end{center}
\caption{Location of the roots of the BAE (\ref{BAE}) for the 
spin $\frac{s}{2}=1$ case ($N=12$, $u=-0.05$, $J=1$, $h=0,1,2$), 
which will give the largest eigenvalue 
for ${\widetilde T}_{2}(v)$ at $v=0$. 
The root forms 2-strings at least for $h=0$.}
\label{roots}
\end{figure}

\begin{figure}
\begin{center}
\includegraphics[width=1\textwidth]
{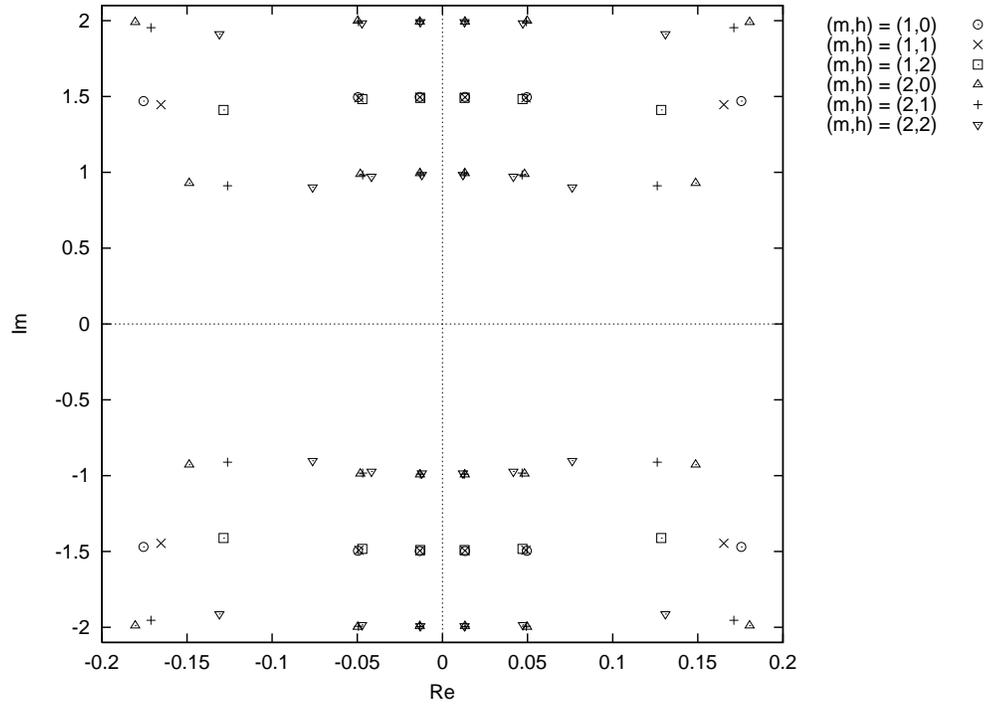}
\end{center}
\caption{Location of the zeros of ${\widetilde T}_{m}(v)$ 
($m=1,2$) for 
the roots in Figure \ref{roots}.  
The zeros are located outside of the physical strip
 ${\rm Im} v \in [-\frac{1}{2},\frac{1}{2}]$.
 }
\label{zeros}
\end{figure}

Following Suzuki's calculations for the spin $\frac{s}{2}=1$ case, 
we have plotted roots of the BAE (\ref{BAE}) in the sector 
$M=N$ and zeros of ${\widetilde T}_{m}(v)$ ($m=1,2$) 
(Figure \ref{roots} and Figure \ref{zeros}). 
Admitting Suzuki's observation, we shall derive the NLIE for 
the largest eigenvalue sector of $\widetilde{ T}_{s}(v)$ from now on. 
$\widetilde{ T}_{m}(v)$ has poles
\footnote{Note that these poles are known ones, which 
we need not solve the BAE (\ref{BAE}) to get.}
 at 
$v=\pm \tilde{\beta}_{m}: \tilde{\beta}_{m}=
(\frac{m+s}{2}+u)i, (\frac{m+s-2}{2}+u)i, 
\dots , (\frac{m+s+2-2\min(m,s)}{2}+u)i$, 
whose order is at most $\frac{N}{2}$. 
Moreover 
\begin{eqnarray}
Q_{m}:=
\lim_{|v| \to \infty} 
\widetilde{ T}_{m}(v)
=\sum_{k=0}^{m}e^{\frac{(m-k)\mu_{1}+k\mu_{2}}{T}}
=\sum_{k=0}^{m}e^{\frac{h(m-2k)}{T}}
\end{eqnarray}
 is a finite number. 
This is a solution of the $Q$-system \cite{K89,KR90}
\begin{eqnarray}
(Q_{m})^{2}=Q_{m-1}Q_{m+1}+Q^{(2)}_{m}, 
\quad m \in {\mathbb Z}_{\ge 1},
\end{eqnarray}
where $Q^{(2)}_{m}=e^{\frac{m(\mu_{1}+\mu_{2})}{T}}=1$ and 
$Q_{0}=1$.
So we must put 
\begin{eqnarray}
\widetilde{ T}_{1}(v)=Q_{1}+\sum_{j=1}^{\frac{N}{2}}
\left\{ \frac{b_{j}}{(v-\tilde{\beta}_{1})^j}+
\frac{\overline{b}_{j}}{(v+\tilde{\beta}_{1})^j} \right\},
\label{expan}
\end{eqnarray}
where $\tilde{\beta}_{1}=(\frac{1+s}{2}+u)i$. 
Here the coefficients are given as follows.
\begin{eqnarray}
b_{j}=\oint_{C} \frac{{\mathrm d} v}{2\pi i}
 \widetilde{T}_{1}(v)(v-\tilde{\beta}_{1})^{j-1},
 \quad 
 \overline{b}_{j}=\oint_{\overline{C}} \frac{{\mathrm d} v}{2\pi i}
 \widetilde{T}_{1}(v)(v+\tilde{\beta}_{1})^{j-1}, 
 \label{coeff}
\end{eqnarray}
where the contour $C$ (resp. $\overline{C}$) is a counterclockwise 
closed loop around $\tilde{\beta}_{1}$ (resp. $-\tilde{\beta}_{1}$) 
that does not surround $-\tilde{\beta}_{1}$ (resp. $\tilde{\beta}_{1}$). 
Substituting (\ref{coeff}) into (\ref{expan}), 
and using (\ref{t-sys2}) and (\ref{t-sys3}), we obtain
\begin{eqnarray}
 \hspace{-15pt}
\widetilde{T}_{1}(v)=Q_{1} &+&
\oint_{C} \frac{{\mathrm d} y}{2\pi i} 
\frac{1-\left(\frac{y}{v-\tilde{\beta}_{1}}\right)^{\frac{N}{2}}}
 {v-y-\tilde{\beta}_{1}}
 \bigg\{ 
 \frac{\widetilde{T}_{m}(y+\tilde{\beta}_{1}-\frac{m-1}{2}i)}
 {\widetilde{T}_{m-1}(y+\tilde{\beta}_{1}-\frac{m}{2}i)} \nonumber \\
 && \hspace{70pt} +
 \frac{\widetilde{g}_{1}(y+\tilde{\beta}_{1}-\frac{i}{2}) 
 \widetilde{T}_{m-2}(y+\tilde{\beta}_{1}-\frac{m+1}{2}i)}
 {\widetilde{T}_{m-1}(y+\tilde{\beta}_{1}-\frac{m}{2}i)}
 \bigg\} \nonumber \\
&+&
\oint_{\overline{C}} \frac{{\mathrm d} y}{2\pi i} 
\frac{1-\left(\frac{y}{v+\tilde{\beta}_{1}}\right)^{\frac{N}{2}}}
 {v-y+\tilde{\beta}_{1}}
 \bigg\{ 
 \frac{\widetilde{T}_{m}(y-\tilde{\beta}_{1}+\frac{m-1}{2}i)}
 {\widetilde{T}_{m-1}(y-\tilde{\beta}_{1}+\frac{m}{2}i)} \nonumber \\
 && \hspace{70pt} +
 \frac{\widetilde{g}_{1}(y-\tilde{\beta}_{1}+\frac{i}{2}) 
 \widetilde{T}_{m-2}(y-\tilde{\beta}_{1}+\frac{m+1}{2}i)}
 {\widetilde{T}_{m-1}(y-\tilde{\beta}_{1}+\frac{m}{2}i)}
 \bigg\} \label{nlie1},
\end{eqnarray}
where the contour $C$ (resp. $\overline{C}$) is a counterclockwise 
closed loop around 0 
that does not surround $-2\tilde{\beta}_{1}$ (resp. $2\tilde{\beta}_{1}$). 
The first term and the second term in the first bracket $\{\cdots \}$ 
in (\ref{nlie1}) have a common pole at $0$. 
However this common pole of the first term disappears if 
$m \ge s+1$. Therefore, for $m \ge s+1$ the first term vanishes after the 
integration as long as the contour $C$ does not surround 
the pole at $\tilde{z}_{m-1}-\tilde{\beta}_{1}+\frac{m}{2}i$. 
Similarly, for $m \ge s+1$, 
the first term in the second  bracket $\{\cdots \}$
vanishes after the integration as long as the contour $\overline{C}$ 
does not surround 
the pole at $\tilde{z}_{m-1}+\tilde{\beta}_{1}-\frac{m}{2}i$. 
Thus, for $m \ge s+1$, we obtain 
\begin{eqnarray}
 \hspace{-15pt}
\widetilde{T}_{1}(v)=Q_{1} &+&
\oint_{C} \frac{{\mathrm d} y}{2\pi i} 
\frac{\widetilde{g}_{1}(y+\tilde{\beta}_{1}-\frac{i}{2}) 
 \widetilde{T}_{m-2}(y+\tilde{\beta}_{1}-\frac{m+1}{2}i)}
{(v-y-\tilde{\beta}_{1})\widetilde{T}_{m-1}(y+\tilde{\beta}_{1}-\frac{m}{2}i)}
  \nonumber \\
&+&
\oint_{\overline{C}} \frac{{\mathrm d} y}{2\pi i} 
 \frac{\widetilde{g}_{1}(y-\tilde{\beta}_{1}+\frac{i}{2}) 
 \widetilde{T}_{m-2}(y-\tilde{\beta}_{1}+\frac{m+1}{2}i)}
{(v-y+\tilde{\beta}_{1})\widetilde{T}_{m-1}(y-\tilde{\beta}_{1}+\frac{m}{2}i)}.
\end{eqnarray}
Here we omit the terms which contain $y^{\frac{N}{2}}$ 
since these terms cancel the poles of $\tilde{g}_{1}$. 
We can take the Trotter limit $N \to \infty$. 
\begin{eqnarray}
 \hspace{-15pt}
{\mathcal T}_{1}(v)=Q_{1} &+&
\oint_{C} \frac{{\mathrm d} y}{2\pi i} 
\frac{g_{1}(y+\beta_{1}-\frac{i}{2}) 
 {\mathcal T}_{m-2}(y+\beta_{1}-\frac{m+1}{2}i)}
{(v-y-\beta_{1}){\mathcal T}_{m-1}(y+\beta_{1}-\frac{m}{2}i)}
  \nonumber \\
&+&
\oint_{\overline{C}} \frac{{\mathrm d} y}{2\pi i} 
 \frac{g_{1}(y-\beta_{1}+\frac{i}{2}) 
 {\mathcal T}_{m-2}(y-\beta_{1}+\frac{m+1}{2}i)}
{(v-y+\beta_{1}){\mathcal T}_{m-1}(y-\beta_{1}+\frac{m}{2}i)},
\label{nlie3}
\end{eqnarray}
where $\beta_{1}=\lim_{N \to \infty} \tilde{\beta}_{1}=\frac{1+s}{2}i$, 
${\mathcal T}_{m}(v)=\lim_{N \to \infty} \widetilde{T}_{m}(v)$ 
and 
\begin{eqnarray}
g_{1}(v)=\exp\left(
\frac{1}{T}\left\{
\frac{Js}{(v^2+\frac{s^2}{4})}
+\mu_{1}+\mu_{2}
\right\}
\right)
=
Q^{(2)}_{1}
\exp\left(
\frac{Js}{(v^2+\frac{s^2}{4})T}
\right).
\end{eqnarray}
In particular, (\ref{nlie3}) for $m=s+1$ is the simplest.
\begin{eqnarray}
 \hspace{-15pt}
{\mathcal T}_{1}(v)=Q_{1} &+&
\oint_{C} \frac{{\mathrm d} y}{2\pi i} 
\frac{g_{1}(y+\frac{s}{2}i) 
 {\mathcal T}_{s-1}(y-\frac{i}{2})}
{(v-y-\frac{s+1}{2}i){\mathcal T}_{s}(y)}
  \nonumber \\
&+&
\oint_{\overline{C}} \frac{{\mathrm d} y}{2\pi i} 
 \frac{g_{1}(y-\frac{s}{2}i) 
 {\mathcal T}_{s-1}(y+\frac{i}{2})}
{(v-y+\frac{s+1}{2}i){\mathcal T}_{s}(y)},
\label{nlie4}
\end{eqnarray}
where the contour $C$ (resp. $\overline{C}$) is a counterclockwise 
closed loop around 0 
that does not surround $-(1+s)i$ (resp. $(1+s)i$) and $z_{s}$. 
(\ref{nlie4}) contains only one unknown function ${\mathcal T}_{1}(v)$ 
since ${\mathcal T}_{s}(v)$ and ${\mathcal T}_{s-1}(v)$ 
can be expressed by ${\mathcal T}_{1}(v)$ 
through (\ref{BR}) in the Trotter limit. 
The free energy per site is given by 
$f=-T \log {\mathcal T}_{s}(0)$. 
For $s=1$, (\ref{nlie4}) reduces to the Takahashi's NLIE for 
the XXX spin chain \cite{Ta01}.
Although we only consider the largest eigenvalue of 
$\widetilde{ T}_{s}(v)$ in the limit $N\to \infty$
, other eigenvalues also satisfy the NLIE (\ref{nlie4}) 
if above conditions for the integral contours 
are satisfied. In usual, such eigenvalues have zeros in 
the physical strip ${\mathrm Im}v \in [-\frac{1}{2},\frac{1}{2}]$. 
Thus to exclude the eigenvalues other than the largest one, 
one may take the integral contours on the line 
${\mathrm Im}v =\pm \frac{1}{2}$. 
\section{High temperature expansion}
In this section, we shall calculate the 
high temperature expansion of the free energy from our 
new NLIE (\ref{nlie4}). 
The calculation is not easier than $s=1$ case \cite{ShT02} 
due to the determinants in (\ref{nlie4}). 
However it is easier than to use the traditional TBA equation. 
For small $J/T$, 
we shall put
\begin{eqnarray}
{\mathcal T}_{m}(v)=\exp
\left(
 \sum_{n=0}^{\infty}b_{m,n}(v)(\frac{J}{T})^n
 \right), \label{hi-expan}
\end{eqnarray}
where $b_{m,0}(v)=\log Q_{m}$. 
Due to (\ref{BR}) in the limit $N\to \infty$, 
one can express $b_{m,n}(v)$ 
in terms of fundamental ones $\{ b_{1,k}(v)\}_{0\le k \le n}$, $Q^{(2)}_{1}$  
and $b(v)=s/(v^2+\frac{s^2}{4})$. 
For example, we have 
\begin{eqnarray}
b_{2,1}(v)=
\left(
 (Q_{1})^2b_{1,1}(v-\frac{i}{2})+
 (Q_{1})^2b_{1,1}(v+\frac{i}{2})-Q^{(2)}_{1}b(v)
 \right)
 /Q_{2} ,\label{depen}
\end{eqnarray}
where $Q_{2}=(Q_{1})^2-Q_{1}^{(2)}$. 
Taking note on the relations like (\ref{depen}), substitute (\ref{hi-expan}) 
into (\ref{nlie4}), we obtain the coefficients $\{b_{m,n}(v)\}$. 
For example, $\{b_{1,n}(v) \}$ for $s=2,n=1,2,3$ are as follows. 
\begin{eqnarray}
&& \hspace{-20pt} b_{1,1}(v)=
\frac{12Q^{(2)}_{1}}
  {\left( 9 + 4v^2 \right) \left( (Q_{1})^2 - Q^{(2)}_{1} \right) },
\nonumber \\
&& \hspace{-20pt} b_{1,2}(v)=
2Q^{(2)}_{1}\Bigl\{ 
\left( 45 + 4v^2 \right) (Q_{1})^4 + 
      \left( -99 + 4v^2 \right) (Q_{1})^2Q^{(2)}_{1} 
      \nonumber \\ 
      && - 
      4\left( 27 + 20v^2 \right) (Q^{(2)}_{1})^2 
      \Bigl\}
     /\left\{
    {\left( 9 + 4v^2 \right) }^2
    {\left( (Q_{1})^2 - Q^{(2)}_{1} \right) }^3
    \right\},
\nonumber \\
&& \hspace{-20pt} b_{1,3}(v)=
\Big\{
Q^{(2)}_{1}
\Bigl( 
\left( 477 + 120v^2 + 16v^4 \right) 
       (Q_{1})^8 + 9\left( -179 + 56v^2 + 16v^4 \right) 
       (Q_{1})^6Q^{(2)}_{1} 
       \nonumber \\ 
&&  - 6
       \left( 1101 + 1536v^2 + 304v^4 \right) (Q_{1})^4
       (Q^{(2)}_{1})^2 
       \nonumber \\ 
&&
+ 4\left( 5553 + 5028v^2 + 896v^4 \right) 
       (Q_{1})^2(Q^{(2)}_{1})^3 
        + 
      48\left( 63 + 84v^2 + 32v^4 \right) (Q^{(2)}_{1})^4 
      \Bigl) 
    \Big\}
\nonumber \\ 
&&
    /\left\{
    {\left( 9 + 4v^2 \right) }^3
    {\left( (Q_{1})^2 - Q^{(2)}_{1} \right) }^5
    \right\}.
\end{eqnarray}
%
We can calculate the specific heat $C=-T \frac{\partial^2 f }{\partial T^2}$
 and the magnetic susceptibility 
 $\chi=- \frac{\partial^2 f }{\partial h^2}|_{h=0}$. 
In this case, we only use $\{b_{m,k}(v)\}$ for $v=0$ due to the definition 
of the free energy (\ref{free}). 
Note that the $h$-dependence of the free energy enters 
only through $Q_{1}=e^{\frac{h}{T}}+e^{-\frac{h}{T}}$ 
since $Q^{(2)}_{1}=1$. 
Let us put $t=J/T$. \\
$s=2$ case:
\begin{eqnarray}
\hspace{-10pt} && \hspace{-10pt} C=
\frac{8\,t^2}{9} + \frac{34\,t^3}{27} - \frac{5\,t^4}{54} - 
  \frac{580\,t^5}{243} - \frac{27629\,t^6}{11664} + 
  \frac{165529\,t^7}{116640} + \frac{40875277\,t^8}{8398080} 
\nonumber \\ \hspace{-10pt} && + 
  \frac{10648871\,t^9}{4408992} - \frac{2176205977\,t^{10}}{470292480} - 
  \frac{94355582827\,t^{11}}{12697896960} - 
  \frac{100181936647\,t^{12}}{304749527040} 
\nonumber \\ \hspace{-10pt} &&  + 
  \frac{23030455724107\,t^{13}}{2394460569600} + 
  \frac{4273471680238097\,t^{14}}{482723250831360} - 
  \frac{919571868309546869\,t^{15}}{188262067824230400} 
  \nonumber \\ \hspace{-10pt} && - 
  \frac{30226224111769481353\,t^{16}}{1916850145119436800} - 
  \frac{1741566568691332074437\,t^{17}}{237210205458530304000} 
  \nonumber \\ \hspace{-10pt} && + 
  \frac{636000568233936625686389\,t^{18}}{45544359448037818368000} + 
  \frac{5325661218179817974957\,t^{19}}{248158368787385548800} 
  \nonumber \\ \hspace{-10pt} && + 
  \frac{143978522302549633307591\,t^{20}}{227535901895503223193600} - 
  \frac{3868537325098539118572347677\,t^{21}}{144433584998668295995392000}
  \nonumber \\
 \hspace{-10pt} &&  +O(t^{22}),
\end{eqnarray}
\begin{eqnarray}
\hspace{-10pt} && \hspace{-10pt} \chi T=
\frac{8}{3} - \frac{8\,t}{3} + \frac{14\,t^3}{27} + 
\frac{205\,t^4}{324} + 
  \frac{97\,t^5}{810} - \frac{32627\,t^6}{58320} - 
  \frac{290839\,t^7}{489888}  + 
  \frac{2993083\,t^8}{26127360} 
  \nonumber \\ \hspace{-10pt} &&
  + \frac{51476713\,t^9}{70543872} + 
  \frac{392473169\,t^{10}}{846526464} - 
  \frac{6147244063\,t^{11}}{14549673600} - 
  \frac{28552400626009\,t^{12}}{33522447974400} 
  \nonumber \\ \hspace{-10pt} &&- 
  \frac{41436528439217\,t^{13}}{186767924428800} + 
  \frac{46001003925515\,t^{14}}{59765735817216} 
   + 
  \frac{5761065533476745581\,t^{15}}{6589172373848064000}
  \nonumber \\ \hspace{-10pt} && 
  - 
  \frac{393055654682062161341\,t^{16}}{2530242191557656576000} - 
  \frac{1142327068285920573167\,t^{17}}{1024145648963813376000}
    \nonumber \\ \hspace{-10pt} && +O(t^{18}).
\end{eqnarray}
$s=3$ case:
\begin{eqnarray}
\hspace{-10pt} && \hspace{-10pt} C=
\frac{15\,t^2}{16} + \frac{145\,t^3}{96} + \frac{1385\,t^4}{2304} - 
  \frac{13445\,t^5}{6912} - \frac{1203755\,t^6}{331776} - 
  \frac{7458199\,t^7}{5971968} 
\nonumber \\ \hspace{-10pt} &&  + 
\frac{66731599\,t^8}{15925248} + 
  \frac{29819652545\,t^9}{4514807808} + 
  \frac{548487273433\,t^{10}}{433421549568} - 
  \frac{62192420939825\,t^{11}}{7801587892224}
  \nonumber \\ \hspace{-10pt} && - 
  \frac{29359344590299711\,t^{12}}{2808571641200640} - 
  \frac{1535815314115199\,t^{13}}{4413469721886720} 
  \nonumber \\ \hspace{-10pt} &&+ 
  \frac{12355531497035499295\,t^{14}}{889755495932362752} + 
  \frac{5343429626131816107323\,t^{15}}{347004643413621473280}
  +O(t^{16}),
\end{eqnarray}
\begin{eqnarray}
\hspace{-10pt} && \hspace{-10pt} \chi T=
5 - 5\,t + \frac{35\,t^3}{54} + \frac{2725\,t^4}{2592} 
+ \frac{3383\,t^5}{5184} - 
  \frac{173425\,t^6}{373248} - \frac{7164095\,t^7}{5878656} 
\nonumber \\ \hspace{-10pt} && - 
  \frac{9795049\,t^8}{13934592} + \frac{634647695\,t^9}{952342272} + 
  \frac{2120381481515\,t^{10}}{1462797729792} + 
  \frac{13785996387509\,t^{11}}{20113468784640} 
\nonumber \\ \hspace{-10pt} &&  - 
  \frac{33615348883175267\,t^{12}}{34756074059857920} - 
  \frac{37654994203903963\,t^{13}}{21515664894197760}+O(t^{14}).
\end{eqnarray}
$s=4$ case:
\begin{eqnarray}
\hspace{-10pt} && \hspace{-10pt} C=
\frac{24\,t^2}{25} + \frac{619\,t^3}{375} + \frac{49159\,t^4}{45000} - 
  \frac{900239\,t^5}{675000} - \frac{257362861\,t^6}{64800000} 
\nonumber \\ \hspace{-10pt} && - 
  \frac{190318307851\,t^7}{58320000000} + 
  \frac{1061704692647\,t^8}{518400000000} + 
  \frac{33531890711924393\,t^9}{4408992000000000} 
\nonumber \\ \hspace{-10pt} && + 
  \frac{13698973673330960069\,t^{10}}{2116316160000000000}- 
  \frac{4091902458911705383\,t^{11}}{1360488960000000000} 
\nonumber \\ \hspace{-10pt} && - 
  \frac{1767044044952349905869283\,t^{12}}{137137287168000000000000}
+O(t^{13}),
\end{eqnarray}
\begin{eqnarray}
\hspace{-20pt} && \hspace{-10pt} \chi T=
8 - 8\,t + \frac{101\,t^3}{135} + \frac{9337\,t^4}{6480} +
 \frac{209797\,t^5}{162000} - 
  \frac{7464779\,t^6}{116640000} 
  \nonumber \\ \hspace{-20pt} &&- 
  \frac{234988285877\,t^7}{146966400000} - 
  \frac{3029100708947\,t^8}{1741824000000} - 
  \frac{393143129330767\,t^9}{3809369088000000} 
  \nonumber \\ \hspace{-20pt} && + 
  \frac{4303789688178371831\,t^{10}}{2285621452800000000} + 
  \frac{278589255303992797247\,t^{11}}{125709179904000000000}
+O(t^{12}).
\end{eqnarray}
\begin{figure}
\begin{center}
\includegraphics[width=0.95\textwidth]
{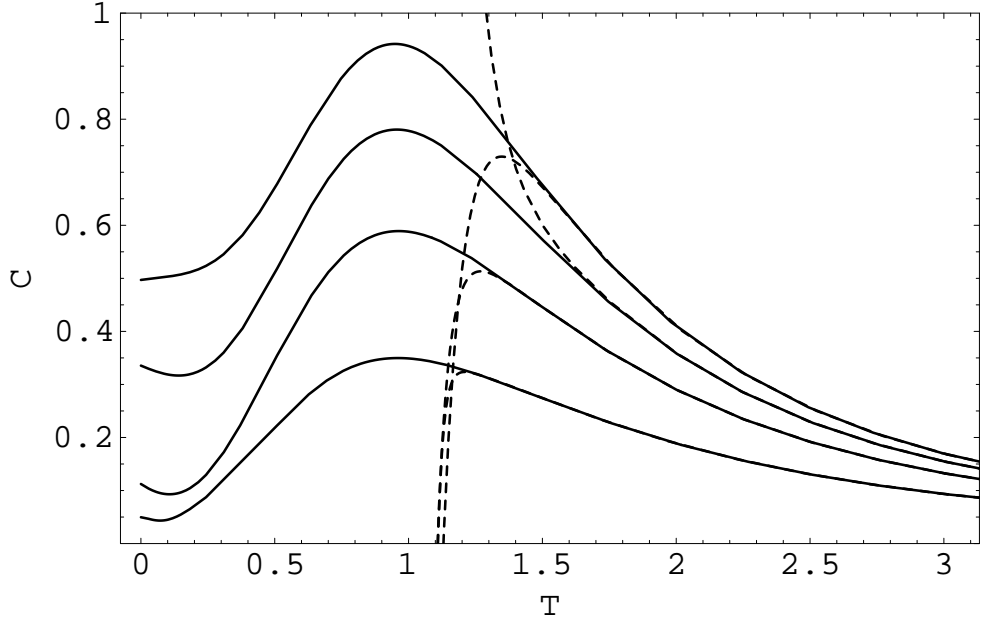}
\end{center}
\caption{Temperature dependence of the high temperature 
expansion of the specific heat $C$ 
for $J=1$ at $h=0$: 
 broken lines denote plain series and 
 smooth lines denote their Pade approximation 
 for $s=1,2,3,4$ 
 from the bottom to the top.  
 Their orders are ($n$: plain series $O(1/T^n)$, Pade)=(52, [26,26]), 
 (21, [10,10]), (15, [7,7]), (12, [6,6]) respectively. 
 Their peak positions and peaks are 
 (peak position, peak)=(0.961, 0.350), (0.963, 0.589), 
 (0.956, 0.780), (0.949, 0.942) respectively. 
 The case for $s=1$ was calculated in ref. \cite{ShT02}.}
\label{specific}
\end{figure}
\begin{figure}
\begin{center}
\includegraphics[width=0.95\textwidth]
{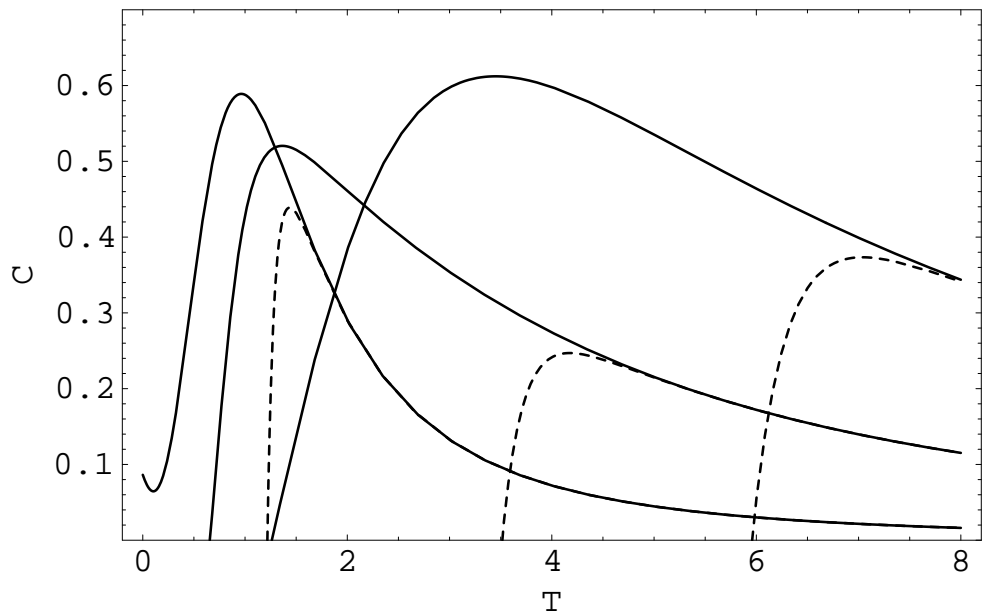}
\end{center}
\caption{Temperature dependence of the high temperature 
expansion of the specific heat $C$ 
for $J=1$, $s=2$: 
 broken lines denote plain series and 
 smooth lines denote their Pade approximation 
 for $h=0,2,4$ 
 from the bottom to the top on the right side.  
 Their order is ($n$: plain series $O(1/T^n)$, Pade)=(17, [8,8]). 
 Their peak positions and peaks are 
 (peak position, peak)= (0.964, 0.589), (1.365, 0.520), 
 (3.452, 0.612) respectively. 
 }
\label{specific2}
\end{figure}
\begin{figure}
\begin{center}
\includegraphics[width=0.95\textwidth]
{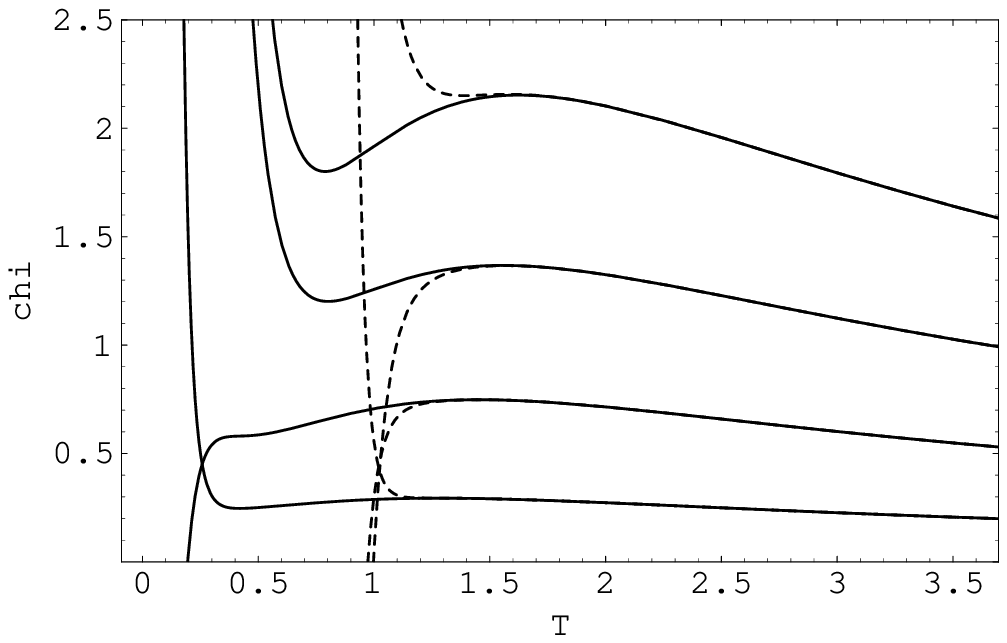}
\end{center}
\caption{Temperature dependence of the high temperature 
expansion of the magnetic susceptibility $\chi$ 
for $J=1$ at $h=0$: 
 broken lines denote plain series and 
 smooth lines denote their Pade approximation 
 for $s=1,2,3,4$ 
 from the bottom to the top on the right side.  
 Their orders are ($n$: plain series $O(1/T^n)$, Pade)=(28, [13,13]), 
 (18, [8,8]), (14, [7,7]), (12, [6,6]) respectively. 
 Their first peak positions from the right and the peaks are 
 (peak position, peak)=(1.282, 0.294), 
 (1.454, 0.748), (1.555, 1.367), (1.624, 2.153) 
 respectively. 
 The case for $s=1$ was calculated in ref. \cite{ShT02}.
 }
\label{magnetic}
\end{figure}
We have plotted 
\footnote{Here we have used the Pade approximation. 
The $[L,M]$ Pade approximation of a function $g(t)$ of $t$ 
is the ratio of a polynomial of degree $L$ and $M$:  
$g(t)=\frac{p_{0}+p_{1}t+p_{2}t^2+\cdots +p_{L}t^L}
{1+q_{1}t+q_{2}t^2+\cdots +q_{M}t^M} +O(t^{L+M+1})$. 
It provides approximately an analytically continued 
function of $g(t)$ for outside of the radius of convergence of $g(t)$ .  
Thus the Pade approximation is expected to provide better 
results for small $T$ than original plain series. 
For more detail, see for example, \cite{DG74}.}
the high temperature expansion of the 
specific heat and the magnetic susceptibility in Figure \ref{specific}-
Figure \ref{magnetic}. 
According to the  Figure \ref{specific}, 
the position of the peak of the specific heat seems not change 
drastically when $s$ changes. 
In particular, $s=2,3$ cases agree with the result 
from another NLIE for large $T$ (see Figure 6 in \cite{S99}). 
This indicates the validity of our new NLIE.  
\section{Concluding remarks}
\noindent
In this paper, we have derived a NLIE with only {\em one} unknown 
function. This type of the NLIE 
for higher spin Heisenberg model with {\em arbitrary} spin 
is derived for the first time. 
In particular, a remarkable 
connection between the NLIE and the quantum Jacobi-Trudi and Giambelli 
formula is firstly found. 

Now that the NLIE is given, the next important task is 
to search the solutions of it. 
The zero-th order of the high temperature expansion (\ref{hi-expan}) 
is a solution of the $Q$-system. 
Thus this task is to incorporate the spectral parameter into 
a solution of the $Q$-system, 
namely to find a solution of the $T$-system. 
The solution of the 
$Q$-system is a kind of a generalization of the 
hypergeometric function (cf. \cite{KNT02}). 
Thus we expect that the final answer will 
be a further generalization of the hypergeometric 
function. 
If a hypergeometric series solution is found, 
one should consider an integral representation of it; then 
one may be able to treat the low temperature 
region where the plain series does not 
converge by an analytic continuation. 

Finally, we note that 
we can also derive a NLIE similar to (\ref{nlie4}) for 
the row to row transfer matrix.
\section*{Acknowledgments}
\noindent
The author would like to thank M. Shiroishi and K. Sakai 
for useful comments on the calculation of the 
high temperature expansion. 
  
\end{document}